\documentstyle[aps,epsf]{revtex}
\begin{document}
\draft

\author{
    E.~M.~Aitala,$^9$
       S.~Amato,$^1$
    J.~C.~Anjos,$^1$
    J.~A.~Appel,$^5$
       D.~Ashery,$^{15}$
       S.~Banerjee,$^5$
       I.~Bediaga,$^1$
       G.~Blaylock,$^8$
    S.~B.~Bracker,$^{16}$
    P.~R.~Burchat,$^{14}$
    R.~A.~Burnstein,$^6$
       T.~Carter,$^5$
 H.~S.~Carvalho,$^{1}$
  N.~K.~Copty,$^{13}$
    L.~M.~Cremaldi,$^9$
 C.~Darling,$^{19}$
       K.~Denisenko,$^5$
       A.~Fernandez,$^{12}$
       P.~Gagnon,$^2$
       K.~Gounder,$^9$
     A.~M.~Halling,$^5$
       G.~Herrera,$^4$
 G.~Hurvits,$^{15}$
       C.~James,$^5$
    P.~A.~Kasper,$^6$
       S.~Kwan,$^5$
    D.~C.~Langs,$^{11}$
       J.~Leslie,$^2$
       B.~Lundberg,$^5$
       S.~MayTal-Beck,$^{15}$
       B.~Meadows,$^3$
 J.~R.~T.~de~Mello~Neto,$^1$
    R.~H.~Milburn,$^{17}$
 J.~M.~de~Miranda,$^1$
       A.~Napier,$^{17}$
       A.~Nguyen,$^7$
  A.~B.~d'Oliveira,$^{3,12}$
       K.~O'Shaughnessy,$^2$
    K.~C.~Peng,$^6$
    L.~P.~Perera,$^3$
    M.~V.~Purohit,$^{13}$
       B.~Quinn,$^9$
       S.~Radeztsky,$^{18}$
       A.~Rafatian,$^9$
    N.~W.~Reay,$^7$
    J.~J.~Reidy,$^9$
    A.~C.~dos Reis,$^1$
    H.~A.~Rubin,$^6$
 A.~K.~S.~Santha,$^3$
 A.~F.~S.~Santoro,$^1$
       A.~J.~Schwartz,$^{11}$
       M.~Sheaff,$^{18}$
    R.~A.~Sidwell,$^7$
    A.~J.~Slaughter,$^{19}$
    M.~D.~Sokoloff,$^3$
       N.~R.~Stanton,$^7$
       K.~Stenson,$^{18}$
    D.~J.~Summers,$^9$
 S.~Takach,$^{19}$
       K.~Thorne,$^5$
    A.~K.~Tripathi,$^{10}$
       S.~Watanabe,$^{18}$
 R.~Weiss-Babai,$^{15}$
       J.~Wiener,$^{11}$
       N.~Witchey,$^7$
       E.~Wolin,$^{19}$
       D.~Yi,$^9$
       S. Yoshida,$^{7}$                         
       R.~Zaliznyak,$^{14}$
       and
       C.~Zhang$^7$ \\
\begin{center} (Fermilab E791 Collaboration) \end{center}
}

\address{
$^1$ Centro Brasileiro de Pesquisas F{\'i}sicas, Rio de Janeiro, Brazil\\
$^2$ University of California, Santa Cruz, California 95064\\
$^3$ University of Cincinnati, Cincinnati, Ohio 45221\\
$^4$ CINVESTAV, Mexico\\
$^5$ Fermilab, Batavia, Illinois 60510\\
$^6$ Illinois Institute of Technology, Chicago, Illinois 60616\\
$^7$ Kansas State University, Manhattan, Kansas 66506\\
$^8$ University of Massachusetts, Amherst, Massachusetts 01003\\
$^9$ University of Mississippi, University, Mississippi 38677\\
$^{10}$ The Ohio State University, Columbus, Ohio 43210\\
$^{11}$ Princeton University, Princeton, New Jersey 08544\\
$^{12}$ Universidad Autonoma de Puebla, Mexico\\
$^{13}$ University of South Carolina, Columbia, South Carolina 29208\\
$^{14}$ Stanford University, Stanford, California 94305\\
$^{15}$ Tel Aviv University, Tel Aviv, Israel\\
$^{16}$ 317 Belsize Drive, Toronto, Canada\\
$^{17}$ Tufts University, Medford, Massachusetts 02155\\
$^{18}$ University of Wisconsin, Madison, Wisconsin 53706\\
$^{19}$ Yale University, New Haven, Connecticut 06511
}
\title{ 
 The doubly Cabibbo-suppressed decay $D^+\to K^+ \pi^- \pi^+$}

\date{\today}

\maketitle


\begin{abstract}
We report  the observation of the doubly Cabibbo-suppressed decay
$D^+\to K^+ \pi^- \pi^+$ in  data from Fermilab charm hadroproduction
experiment E791. With a signal of 59 $\pm$ 13 events we measured the ratio of
the branching fraction for this mode to that of the Cabibbo-favored decay
$D^+\to K^- \pi^+ \pi^+$ to be $B(D^+ \to K^+ \pi^- \pi^+)$ /
$B(D^+ \to K^- \pi^+ \pi^+$) $=$
($7.7 \pm 1.7 \pm 0.8) \times 10^{-3}$. A Dalitz plot analysis was
performed to search for resonant structures.
\end{abstract}

\narrowtext

~\\
The origin of the differences between the charm meson
lifetimes is associated with their hadronic decays. While the
semileptonic decay rates of the $D^0$ and $D^+$ are the same, the 
Cabibbo-favored (CF) hadronic decay rate of the $D^0$ is 3.2 times
that of the 
$D^+$. There are at least two possible sources for this difference.
The CF $D^+$ hadronic decay rate could be suppressed by destructive 
interference between spectator amplitudes containing indistinguishable 
final state quarks. It is also possible that
the CF $D^0$ decay rate is  enhanced by 
non-spectator amplitudes which do not exist for the $D^+$.

For both hadronic CF $D^0$ decays and doubly Cabibbo-suppressed (DCS) $D^+$
decays, all the final state quarks have different flavors, thus removing the
possibility of destructive interference. In the simplest picture,
non-spectator amplitudes are small enough to ignore, and one would expect
$\Gamma_{DCS} (D^+)/ \Gamma_{CF} (D^0) \approx \tan^4 \theta_C $ and
$\Gamma_{DCS} (D^+)/ \Gamma_{CF} (D^+) \approx 3.2 \times \tan^4 \theta_c $.
These relations need not be satisfied if non-spectator amplitudes are also
important. Doubly Cabibbo-suppressed  decays can thus provide important
insights into the  $D$ meson lifetime pattern.

In this paper we report a measurement from Fermilab experiment E791
of the branching fraction for the DCS decay $D^+ \to K^+ \pi^- \pi^+$.
Throughout this paper, reference to $D^+$ and $D_s^+$ and to their decay 
modes imply also the corresponding charge-conjugate states.

The data were recorded from 500 GeV/$c$~ $ \pi^- $ interactions in five thin
foils (one platinum, four diamond) separated by gaps of 1.34 to 1.39~cm. 
The experiment recorded $ 2 \times 10^{10} $ events with a loose transverse
energy trigger.

The E791 spectrometer was an upgraded version of the apparatus used in 
Fermilab experiments E516, E691, and E769~\cite{OLDTPL}. 
Position information for track and vertex reconstruction was provided by 
23 silicon microstrip detectors (6 upstream of the target foils,
17 downstream), 10 proportional wire chamber planes (8 upstream and
2 downstream of the target) and 35 drift chamber planes.
Momentum analysis was provided by two dipole magnets which bent particles 
in the horizontal plane.
Particle identification was performed by two segmented threshold 
\v{C}erenkov
counters~\cite{CKV}, allowing unambiguous
identification of pions and kaons in the 
momentum range from 6 to 40 GeV/$c$.

After reconstruction, events with evidence of well-separated
production (primary) and decay  (secondary) vertices were retained for further
analysis. 
The position resolutions along and transverse to the beam 
direction for the primary vertex
were $350\,\mu$m and $6\,\mu$m, respectively. For 3-prong secondary
vertices from $D^+$ decays, the transverse resolution was about $9\,\mu$m,
nearly independent of the $D^+$ momentum; the longitudinal resolution was about
360~$\mu$m for a $D^+$ momentum of 70~GeV/$c$ and increased roughly linearly 
with a slope of $30\,\mu$m per 10~GeV/$c$.

We selected a generic $K\pi\pi$ sample containing both DCS 
$D^+ \to K^+ \pi^- \pi^+$ and  CF $D^+ \to K^- \pi^+ \pi^+$ decay
candidates. The abundant CF decay  was used to determine 
the track and vertex selection criteria used in the 
search for the DCS decay.  
The criteria were chosen to maximize $N_S/\sqrt{N_B}$, 
where $N_S$ and $N_B$ are the numbers of signal
and background events in the $K^- \pi^+ \pi^+$ sample.

We required the secondary vertex to be well-separated from the primary vertex
and located well outside the target foils and other solid material, the
momentum vector of the candidate $D^+$  to point back to the primary vertex,
and the decay track candidates to pass closer to the secondary vertex than to
the primary vertex. We used longitudinal separations normalized by their
resolutions to reduce momentum-dependent effects. Specifically, a
3-prong secondary vertex had to be separated by at least 20 $\sigma_L$ from the
primary vertex and by at least 5 $\sigma_L$ from the closest material in the
target foils, where the $\sigma_L$ are 
resolutions in the measured longitudinal separations.
The sum of the momentum vectors of the three
tracks from this secondary vertex could not miss the primary vertex by more
than 40$\mu$m in the plane perpendicular to the beam. We formed the ratio of
each track's smallest distance from the secondary vertex to its smallest
distance from the primary vertex, and required the product of these ratios for
the three tracks to be less than 0.001.

In addition to the selection criteria described above, we required \v{C}erenkov
particle identification for all three  decay candidate tracks. The 
\v{C}erenkov efficiencies
and corresponding misidentification rates were measured using the CF signal, in
which the particle identification of the decay tracks can be determined from
their charge. In this analysis the efficiency for correctly identifying kaons
was 45\%; the corresponding probability of misidentifiying real pions as kaons
was 2\%. Since misidentification of the odd-charged pion candidate was a large
source of contamination, we used a more stringent identification criterion for
this track than that for the like-charged pion. For the odd-charged pion, the
efficiency for correct identification was 57\%, and the probability of
misidentifying  kaons as pions was  13\%. For the like-charged pion the
efficiency for correct identification was 85\%; the corresponding probability
of misidentifying kaons as pions was  37\%. The overall particle identification
efficiency for $ D^+ \rightarrow K \pi \pi $ was  22\%.

Due to particle misidentification and reconstruction errors, several other
charm decays contributed to the  background in the $K^+ \pi^- \pi^+$ sample.
The major sources of charm background are listed below.

a) $D^+$ and $D_s$ decays with missing neutrals, such as
$D^+ \to ~\bar K^{*0} l^+ \nu$, $D^+ \to K^- \pi^+ \pi^+ \pi^0$
and $D_s \to \phi l^+ \nu$. 
In these cases 4-body decays produced 3-prong vertices. 
Monte Carlo simulations showed that, 
because of particle misidentification,  these events
are spread smoothly across the entire $K^+ \pi^- \pi^+$ mass spectrum.
Their contribution was included with those of type (b) in a smooth background
whose level was determined from the fit discussed below.

b) $D^0$ decays such as $D^0 \to K\pi$ and $D^0 \to K\pi\pi\pi$. 
Such events passed the selection criteria for $K\pi\pi$ when the reconstruction
algorithm found two tracks from such a charm decay and combined them with a
third track to form a spurious 3-prong vertex, or when one track was lost from
a 4-body decay.
False vertices created from $D^0 \to K^- \pi^+$ plus another track populate the
$K^+ \pi^- \pi^+$ mass spectrum above 2 GeV/$c^2$, and were eliminated by
explicitly removing events with $K^{\pm} \pi^{\mp}$ invariant mass between
1.828  GeV/$c^2$ and 1.900 GeV/$c^2$. A large fraction of the charm background
originated from $ D^0 \to K^- \pi^- \pi^+ \pi^+ $ decays where one of the
like-charged pions was lost and the remaining tracks were correctly identified.
Since these events were concentrated below 1.76 GeV/$c^2$ the fit was
restricted to the region above 1.76 GeV/$c^2$. Monte Carlo simulations
indicated that the remaining events from the $ D^0 \to K^- \pi^- \pi^+ \pi^+ $
background were  smoothly spread across the $K^+ \pi^- \pi^+$ mass spectrum.

c) $D^+$ and $D_s$ 3-body hadronic decays. 
These were the most problematic backgrounds because they produce structures in
the $K^+ \pi^- \pi^+$ mass distribution. Here, the 3-prong candidates came from
real 3-body decays whose reflections in the $K^+ \pi^- \pi^+$ mass spectrum
were concentrated at shifted masses, except for the CF $ D^+ \rightarrow K^-
\pi^+ \pi^+ $ decay whose reflection was smoothly spread across the $K^+ \pi^-
\pi^+$ mass distribution. The  $ D_s $ and $ D^+ \rightarrow KK\pi$ final
states had very clean $\phi \pi$ components. We therefore eliminated events
with $K^+K^-$ invariant mass in the range from 1.005  GeV/$c^2$ to 1.035
GeV/$c^2$. This removed 2\% of true CF and DCS  decays.

The range of $K^+ \pi^- \pi^+$ masses over which we could
reliably model the charm background was 1.76 to 2.06 GeV/$c^2$. Within
this interval, backgrounds of type a) and b) did not produce peaks. The
structures resulting from background c) are shown in Figure 1. The parameters
for the reflection shapes were determined by intentionally misidentifying
tracks in the background channels. This was done with real data for $ D^+ \to
K^- \pi^+ \pi^+ $ and $ D^+ , D_s \to K K \pi $, and with Monte Carlo events
for $ D^+ \to \pi \pi \pi $. The net effect of these reflections is to produce
a small $K^+\pi^-\pi^+$ enhancement in the vicinity of the $D^+$ mass.

\begin{figure}
\centerline{\epsfysize=3.00in \epsffile{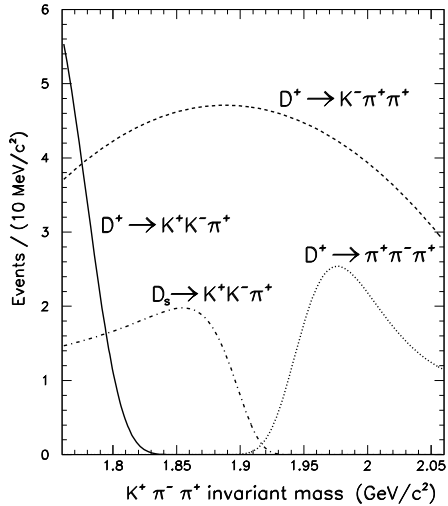}}
\caption{
Distributions of the hadronic 3-body charm decay backgrounds
that produce structures in the $K^+ \pi^- \pi^+$ mass plot. The area under
each curve represents the estimated number of background events of each type.}
\end{figure} 

The area under each curve in Figure 1 represents the estimated number of
background events of each type, as fixed in the final fit. The amount of each
background was determined by a rapidly-converging iterative procedure. The
candidate events from the $K^+ \pi^- \pi^+$ sample were successively plotted
as though they were $K^+ K^- \pi^+$, $\pi^+\pi^+\pi^-$ or $K^- \pi^+\pi^+$,
with an increasingly accurate description of the feedthrough from the other
channels. The result of this procedure was that in the $K^+ \pi^- \pi^+$ mass
range between 1.76 and 2.06 GeV/$c^2$ there are
125 $\pm$ 13 $ D^+ \rightarrow K^- \pi^+ \pi^+ $ events, 
25 $\pm$ 10 $D^+ \to \pi^+\pi^+\pi^-$ events, 
24 $\pm$ 11 $D_s \to K^+ K^-\pi^+$ events,
and 15 $\pm$ 4 $D^+ \to K^+ K^-\pi^+$ events. 

The $K^+ \pi^- \pi^+$ mass distribution for the
final sample of  decay candidates is shown in
Figure 2. The spectrum was fit to the sum of the reflections described above, a
smooth function which describes the sum of all other backgrounds, a Gaussian
function representing the $ D^+ \to K^+ \pi^- \pi^+ $ signal, and a Gaussian
representing the singly Cabibbo-suppressed $ D_s \to K^+ \pi^- \pi^+ $ signal.
The smooth background was modeled by an exponential function whose parameters
were allowed to vary freely. The centroid and width of the Gaussian describing
the DCS signal were fixed to the values measured for the corresponding CF
signal, 1.870 GeV/$c^2$ and 12 MeV/$c^2$, as described below. The centroid and
width of the $ D_s \to K^+ \pi^- \pi^+ $ signal were fixed at 1.970 GeV/$c^2$
and 12 MeV/$c^2$. The number of $D^+ \to K^+ \pi^- \pi^+$ signal events
determined by the fit was $ 59 \pm 13 $.

\begin{figure}
\centerline{\epsfysize=3.00in \epsffile{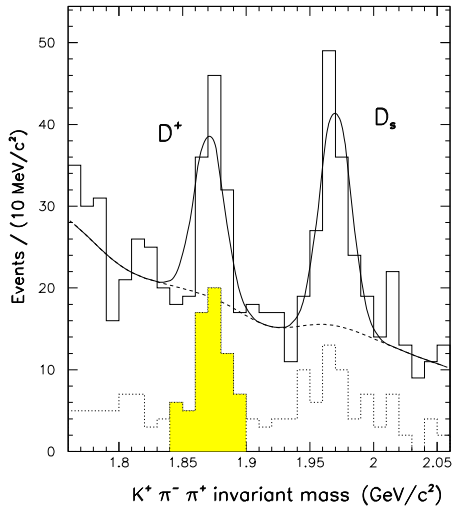}}
\caption{Mass spectrum of candidate $K^+ \pi^- \pi^+$ events with 
selection criteria optimized for the branching fraction measurement.
The total background, including the contributions shown in Figure 1,
is represented by the solid line outside the peaks and the dashed line
under the peaks. The
dotted histogram shows the mass distribution with tighter selection criteria.
The events in the shaded area were used for the Dalitz plot analysis.}
\end{figure}

The $K^- \pi^+ \pi^+$ mass distribution
for the sample of CF $D^+ \to K^- \pi^+ \pi^+$ events, selected with the same
criteria as the DCS signal and used as the normalization 
signal, is shown in Figure 3.
The spectrum was fit to the sum of a linear background and a Gaussian
signal whose parameters were allowed to vary, yielding the values quoted
above. The number of $ D^+ \to K^- \pi^+ \pi^+ $ signal events was found to
be $ 7688 \pm 90 $.

Using Monte Carlo simulations with resonant and 
nonresonant decay modes, we found that 
the product of acceptance and efficiency was the same for the CF and DCS 
samples within the $\pm 2\%$ statistical errors in the simulations.
The ratio of branching fractions for the DCS $D^+ \to K^+\pi^-\pi^+$
and CF $ D^+ \to K^- \pi^+ \pi^+ $ decay modes is, thus, given
by the ratio of the measured 
numbers of DCS and CF signal events,

\begin{equation}
{ B(D^+ \to K^+ \pi^- \pi^+)  \over B(D^+ \to K^- \pi^+ \pi^+) } = 
 ( 7.7 \pm 1.7 \pm 0.8 ) \times 10^{-3} .
\end{equation}

The first error reported is statistical. The second is the 
systematic error, which was dominated by uncertainties
in the background shape (8.5\%), in the estimated number of background
events used in the fit (4\%), and the systematic error associated
with particle identification (3.8\%). The total fractional systematic 
error is 10\%.

Using the PDG value \cite{PDG96} for the CF branching fraction, 
$ (9.1 \pm 0.6) $\%, we find 
\begin{equation}
B(D^+ \to K^+ \pi^- \pi^+) =
 (7.0 \pm 1.5 \pm 0.9) \times 10^{-4},
\end{equation}
where the fractional uncertainty in $ B (D^+ \to K^- \pi^+ \pi^+ )$ has been
added in quadrature with our systematic uncertainty in the ratio of branching
fractions to determine the systematic error for the absolute DCS branching
fraction.

The value we have measured for the ratio of DCS to CF branching fractions
is (3.0 $\pm$ 0.8)$\times \tan^4 \theta_C $, which agrees well with the 
simple spectator picture discussed in the introduction.
It also agrees well with the recent result from Fermilab E687\cite{e687}
for ratio (1), $(7.2 \pm 2.3 \pm 1.7)\times 10^{-3}$.

\begin{figure}
\centerline{\epsfysize=3.00in \epsffile{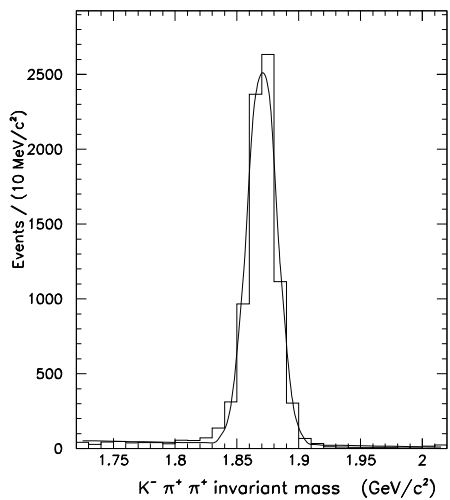}}
\caption{Mass spectrum of candidate $D^+\to K^- \pi^+ \pi^+$ events with final
optimized selection criteria. The curve is a fit to a Gaussian signal and a 
linear background. The DCS $D^+ \to K^+ \pi^- \pi^+$ signal was normalized 
to this signal.}
\end{figure}

To study the DCS amplitudes which lead to the decay $ D^+ \to K^+ \pi^- \pi^+ $,
we have analyzed the Dalitz plot of a smaller but cleaner sample of events.
Because so much of the background in the larger sample comes from misidentified
charm decay, we used even more stringent particle identification on the
odd-charged pion, which reduced the particle identification efficiency from
22\% to 17\%. We also required  that each candidate's proper decay time be
greater than two $D^0$ lifetimes to suppress background from $ D^0 $ and $D_s $
decays. We also explicitly removed candidates consistent with the 
mass hypothesis for $ D^+ \to K^- \pi^+ \pi^+ $. 
This tighter selection gives a branching ratio consistent with that from
the larger sample (equation (1)), for which selection criteria were chosen to
maximize the projected sensitivity.
The use of this cleaner
sample, which contains 42
$\pm$ 9 signal events, reduced the systematic uncertainties from parametrizing
backgrounds in the amplitude analysis.
The Dalitz plot analysis was restricted
to events found within 30 MeV/$c^2$ of the $ D^+ $ mass (67 events altogether).
These events correspond to the shaded area in the histogram at the bottom of
Figure 2. The Dalitz plot of these events is shown in Figure 4.

\begin{figure}
\centerline{\epsfysize=2.90in \epsffile{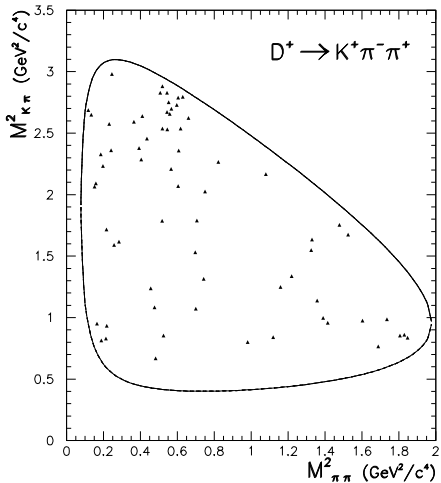}}
\caption{Dalitz plot 
($M^2(K^+ \pi^-)$ vs. $M^2(\pi^+\pi^-)$)
of events found within 30 MeV/$c^2$ of the $ D^+ $ mass.
This sample corresponds to the events in the shaded region of the
dotted histogram shown in Figure 2.}
\end{figure}

The Dalitz plot was fit using the unbinned maximum-likelihood method. 
The $D^+ \to K^+ \pi^- \pi^+$ decay amplitude was represented by a uniform 
nonresonant component
plus relative amplitudes corresponding to the decays 
$D^+ \to K^+ \rho^0(770)$ and $D^+ \to K^{*0}(890) \pi^+$,

\begin{equation}
{\cal M} =  A_{NR}+
\sum_{j=1}^2 a_j e^{i \delta_j} R_j (m_{K\pi}^2,m_{\pi\pi}^2).
\end{equation}

Each resonant component $R_j$ was parameterized by a relativistic Breit-Wigner
function, of constant width, multiplied by a function describing the angular
distribution of decay particles. The nonresonant mode $A_{NR}$ 
was chosen as the
reference channel, fixing the scale for the relative fractions and the phase
convention $(A_{NR}\equiv 1)$.
The fit parameters were, therefore, the relative phases $\delta_j$ and the
real positive coefficients $a_j$ for each resonant amplitude. The signal
likelihood was obtained by multiplying the ideal Dalitz plot density by a
function describing the geometrical acceptance and the reconstruction and
event-selection efficiencies, including the removal of
$\phi \rightarrow K^+ K^-$ and $D^+ \rightarrow K^- \pi^+\pi^+$ events
described above.

The background was represented by the sum of a constant term plus a 
product of two  Gaussians in $K^+ \pi^-$ and $\pi^+ \pi^-$ masses
accounting for the $D_s \to K^+ K^- \pi^+$ 
reflection, which is 
concentrated in the upper part of the Dalitz plot. The shape of the $D_s$
reflection was obtained from Monte Carlo $D_s \to K^+ K^- \pi^+$ events that
passed through the same selection criteria as for the $K^+ \pi^- \pi^+$ sample. We estimate
5 $\pm$ 3 $D_s$ decays in the sample shown in Figure 4.

The fit results are shown in Table 1. The decay fractions  were computed by
integrating the  squared amplitude of each mode over the phase space, and then
dividing it by the integral of the square of the sum of all amplitudes. As
shown in Table 1, the contributions of the three components are comparable. The
two resonant modes are approximately in phase with each other and roughly
$90^\circ$ out of phase with the nonresonant part.

Because of the $D_s$ contribution to the cluster of events at the top of the
Dalitz plot, the fractions corresponding to the $K^+ \rho^0(770)$ and
nonresonant modes are anticorrelated with the estimated number of $D_s \to K^+
K^- \pi^+$ background events. Changing the estimated $D_s$ background to 8
events causes the $K^+ \rho^0(770)$ fraction to decrease by about 
0.5~$\sigma$, where $\sigma$ is the statistical uncertainty from the fit,
and the nonresonant fraction to increase by the same amount. If, instead, the
expected $D_s$ background were 2 events, the $K^+ \rho^0(770)$
fraction would increase and nonresonant fraction decrease by 0.5~$\sigma$.
These uncertainties dominate the systematic error.

We also attempted to include a $D^+ \to K^*(1430) \pi^+$ amplitude in the 
fit, but this broad ($\Gamma =$ 287 MeV) spin zero resonance
could not be distinguished from the nonresonant amplitude.

\begin{table}

\caption{Results from  the $D^+ \to K^+ \pi^- \pi^+$ Dalitz plot fit. There
is an anticorrelated systematic error of $\pm$ 0.07 on the fractions of 
the nonresonant and $K^+\rho^0(770)$ modes, due to uncertainty in the 
estimated number of $D_s \to K^+ K^- \pi^+$ background events.
This uncertainty in $D_s$ background has negligible effect on the phases.
\label{1} }

\begin{tabular}{ccc}   
mode                &   phase (radians) &    fraction         \\ \hline \hline
nonresonant        &    0 (fixed)      & $0.36 \pm 0.14 \pm 0.07$    \\ 
$K^{*0}(890) \pi^+$ & $1.8 \pm 0.5$     & $0.35 \pm 0.14 \pm 0.01$  \\ 
$K^+\rho^0(770)$ & $2.0 \pm 0.4$     & $0.37 \pm 0.14 \pm 0.07$    \\ 
\end{tabular}

\end{table}
\clearpage

The fit results are also shown in Figure 5 as Monte
Carlo density plots. On the left is the generated signal probability 
distribution function (pdf); on the right is the signal  plus background pdf,
now multiplied by efficiency and acceptance functions. 
Due to the more stringent particle-identification criterion adopted for the
odd-charged pion candidate, the overall efficiency was reduced at low values
of the squared masses. This can be seen clearly by comparing plots a) and b)
in Figure 5.

A Monte Carlo technique was used  both to test the fitting procedure and to 
assess the goodness-of-fit. A large number of $K^+ \pi^- \pi^+$ samples 
was generated 
according to the overall probability distribution function, using as input
parameters the phases and coefficients given by the fit to the real data.
Each of these Monte Carlo samples was then fitted, and the distributions of
the resulting fit parameters were plotted.
The average values of all fit parameters were the same as their input values,
and the rms spread in each fit parameter distribution was close to the 
error on this parameter given by the fit of real data. The
fraction of Monte Carlo samples for which the value of 
$-2 \ln{({\cal L}_{max})}$
exceeds that of real data, 
where ${\cal L}_{max}$ is the maximum value of the
sample likelihood, estimates the confidence level of our fit to be 52\%.

\begin{figure}
\centerline{\epsfysize=3.00in \epsffile{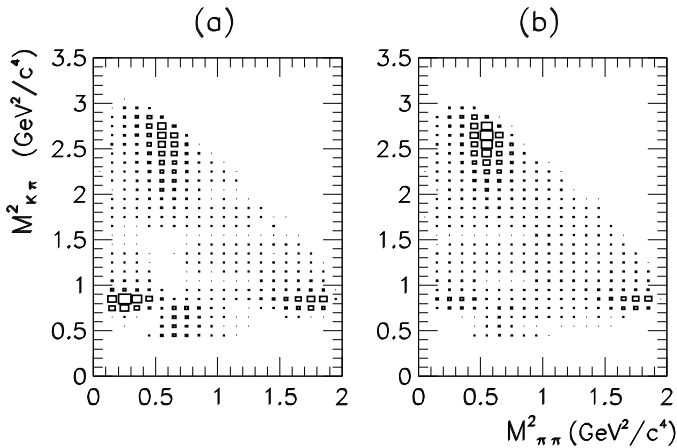}}
\caption{Monte ~Carlo ~simulations ~of ~the ~$ D^+ \to K^+ \pi^- \pi^+$ 
~Dalitz ~plot for
(a) generated signal and (b) accepted signal plus background.
Both use the signal parameters of the fit reported in Table I.
Plot (b) also includes the background measured for
the data of Figure 4. In these plots, the linear dimension of the boxes
is proportional to the number of events.}
\end{figure}

In spite of the large errors on the fractions, the pattern 
shown in the DCS $D^+ \to K^+ \pi^- \pi^+$
decay appears to differ qualitatively  from the corresponding CF decay, 
in which the nonresonant component corresponds to about 95\% of the branching 
fraction \cite{PDG96}.

In summary, Fermilab experiment E791 has measured the 
ratio of branching fractions for the doubly Cabibbo-suppressed 
$D^+ \to K^+\pi^-\pi^+$ and the Cabibbo-favored $ D^+ \to K^- \pi^+ \pi^+ $ 
decay modes, $B(D^+ \to K^+ \pi^- \pi^+)$ /  
$B(D^+ \to K^- \pi^+ \pi^+) = (7.7 \pm 1.7 \pm 0.8) \times 10^{-3}$,
corresponding to  $(3.0 \pm 0.8)\,\tan^4{\theta_C}$. A
Dalitz plot analysis indicates that the DCS signal is composed of
approximately equal amounts of  $D^+ \to K^+ \rho^0(770)$, 
$D^+ \to K^{*0}(890) \pi^+$ and nonresonant modes.
Using the measured fractions from Table 1 and the $D^+ \rightarrow
K^+\pi^-\pi^+$ branching fraction from equation (2), we obtain
$B(D^+\rightarrow K^{*0}(890)\pi^+)= (2.5\pm 1.2) \times 10^{-4}$,
$B(D^+\rightarrow K^+\rho^0(770))=(2.6\pm 1.3)\times 10^{-4}$, and
$B(D^+\rightarrow {\rm{nonresonant~}} K^+\pi^-\pi^+)=
(2.5\pm 1.3)\times 10^{-4}$; statistical and systematic errors have been
added in quadrature.

We gratefully acknowledge the assistance of the staffs of Fermilab and of all
the participating institutions.  This research was supported by the Brazilian
Conselho Nacional de Desenvolvimento Cient\'{\i}fico e Tecnol\'{o}gico,
CONACyT (Mexico), the Israeli Academy of Sciences and Humanities, 
the U.S. Department of Energy, the U.S.-Israel
Binational Science Foundation, and the U.S. National Science Foundation.
Fermilab is operated by the Universities Research Association, Inc., under
contract with the United States Department of Energy.

\small
\bibliographystyle{unsrt}

\begin{thebibliography}{99.}
\small



\bibitem{OLDTPL} 
J.~A. Appel, Ann.~Rev.~Nucl.~Part.~Sci.~42 (1992) 367, and
references therein;
D.~J. Summers {\em{et al.}}, Proceedings of the XXVII$^{\rm{th}}$ Rencontre
de Moriond, Electroweak Interactions and Unified Theories,
Les Arcs, France (15 - 22 March 1992) 417;
S.~Amato {\em{et al.}}, Nucl. Instr. Meth.~A324 (1993) 535.


\bibitem{CKV}
D. Bartlett {\em{et al.}}, Nucl. Instr. Meth.~A260 (1987) 55.




\bibitem{PDG96} 
Particle Data Group, L. Montanet {\em{et al.}}, 
Phys. Rev.~D54 (1996) 1.

\bibitem{e687}
P. L. Frabetti {\em et al.}, Phys. Lett.~B359 (1995) 403.

\end{thebibliography}

\end{document}